\documentclass[twocolumn,prd,amsmath,preprintnumbers,amssymb,superscriptaddress,nofootinbib]{revtex4}
\usepackage{hyperref}
\usepackage{graphicx}
\newcommand {\beq}{\begin{equation}}
\newcommand {\eeq}{\end{equation}}
\newcommand {\beqa}{\begin{eqnarray}}
\newcommand {\eeqa}{\end{eqnarray}}
\newcommand {\n}{\nonumber \\}
\newcommand {\ks}{k \hspace*{-2mm} \slash }

\newcommand {\p}{\partial}
\newcommand {\tr}{{\rm tr}}
\newcommand {\Tr}{{\rm Tr}}
\newcommand {\Str}{{\rm Str}}

\begin{document}

\preprint{KEK-TH-1191}

\title{Superstring vertex operators in type IIB matrix model}

\date{published 20 June 2008}

\author{Yoshihisa Kitazawa}
\email{kitazawa@post.kek.jp}
\affiliation{High Energy Accelerator Research Organization (KEK), 
Tsukuba, Ibaraki 305-0801, Japan}
\affiliation{Department of Particle and Nuclear Physics
The Graduate University for Advanced Studies
Tsukuba Ibaraki 305-0801 Japan}
\author{Satoshi Nagaoka}
\email{nagaoka@post.kek.jp}
\affiliation{High Energy Accelerator Research Organization (KEK),
Tsukuba, Ibaraki 305-0801, Japan}

\begin{abstract}
We clarify the relation between the vertex operators in 
type IIB matrix model and superstring.
Green-Schwarz light-cone closed superstring theory is obtained from 
IIB matrix model on two dimensional noncommutative backgrounds. 
Superstring vertex operators should be reproduced from those of 
IIB matrix model through this connection. 
Indeed, we confirm that supergravity 
vertex operators in IIB matrix model on the two dimensional backgrounds
reduce to those in superstring theory.
Noncommutativity plays an important role in our
identification.
Through this correspondence,
we can reproduce superstring 
scattering amplitudes from IIB matrix model.
\end{abstract}

\maketitle

\section{Introduction}

String theory is perturbatively reproduced from type IIB matrix model
\cite{IKKT} via two dimensional noncommutative (NC) ${\cal N}=8$ 
supersymmetric gauge theory
\cite{CDS,AIIKKT,Li} in the IR limit \cite{KN3}.
The emergent string theory is Green-Schwarz (GS) type IIA superstring theory 
with light-cone gauge \cite{GS4}, 
which is also derived from commutative 
${\cal N}=8$ super Yang-Mills \cite{DVV}.

In the two dimensional gauge theory,
the IR limit corresponds to the strong coupling limit
since the gauge coupling has dimension $(\text{length})^{-1}$.
Dijkgraaf, Verlinde and Verlinde have shown that
the IR limit corresponds to the free string limit.
This logic is applicable to our case and
actually we have derived the free Green-Schwarz string action 
from NC gauge theory by taking the strong coupling limit \cite{KN3}.
We emphasize here that the strong coupling limit (the IR limit)
of NC gauge theory is the free string limit, not the
low energy limit of Green-Schwarz string theory.

The winding number $w$ is reinterpreted as a light-cone momentum $p_+$
in a T-dual interpretation.
In NC gauge theory, such a duality is realized by identifying the 
momentum with the longitudinal coordinates. 
In our procedure, noncommutativity $\theta$ plays a crucial role
to reproduce the worldsheet action.
It is identified with the string scale $\alpha'$ 
in the process of derivation. 

Noncommutativity in NC gauge theory 
gives rise to various novel properties
in comparison with the commutative gauge theory.
Among them, we quote the following two aspects.
The first property is the regularization of UV divergence,
which gives rise to the UV/IR mixing effect \cite{MRS}.
The qualitative 
behavior of the correlation functions, such as the power of the momentum
dependence, is affected by the existence of noncommutativity.
The second property is the introduction of the scale in the theory.
Since there are no scale parameters in the action of IIB matrix model,
the noncommutativity could provide a fundamental scale
in the emergent theory.
We have indeed identified
the scale in NC gauge theory with the string scale 
in the previous paper \cite{KN3}.
Since the NC scale sets the eigenvalue density of the matrices,
our identification of the string scale is consistent with \cite{KS}.

In order to reveal the perturbative superstring picture in 
IIB matrix model more explicitly,
it is important to 
demonstrate the procedure to
calculate the
superstring scattering amplitude from IIB matrix model.
Since there is an open/closed duality in string theory, 
closed string theory will be naturally included in IIB matrix model.
The massless sector of type II closed superstring consists of the 
supergravity multiplet.
Supergravity controls the behavior of long range forces.
The coupling between the fields in the supergravity multiplet and 
the operators in 
IIB matrix model has been clarified through the construction of 
the relevant vertex operators.
The vertex operators for the supergravity multiplet are determined 
uniquely by the maximal ${\cal N}=2$ supersymmetry
transformation in IIB matrix model \cite{vertex,ITU,KMS}. 
These operators are constructed in \cite{vertex} in the first study, 
where the wave functions are introduced as the representations of supergravity 
multiplet. In \cite{ITU}, 
the vertex operators are investigated further
by expanding the supersymmetric Wilson loop
operators \cite{Hamada}.
In \cite{KMS}, the vertex operators are constructed up to the six-th order
of Majorana-Weyl spinor $\lambda$.
Since the algebraic calculation is very complicated, 
the complete structure of the vertex operators is not yet determined,
but the vertex operators for 
the conjugate gravitino and two-form antisymmetric field are
completely determined.

In this paper, 
we compare the IIB matrix model 
vertex operators on the 
two dimensional backgrounds
with those in superstring.
As we have derived supersymmetry transformation of GS light-cone 
superstring from IIB matrix model on the 
two dimensional backgrounds \cite{KN3},
we can reconstruct the superstring vertex operators 
based on the symmetry.
In such a sense, this comparison can be regarded as a
consistency check of the IIB matrix model
vertex operator construction.
Through this correspondence,
we can reproduce
the physical superstring scattering amplitudes from 
IIB matrix model.

In section \ref{31}, we review the supergravity vertex operators
in IIB matrix model, which was 
constructed in \cite{KMS}. 
In section \ref{32}, we review the closed 
superstring vertex operator construction of
GS light-cone superstring theory.
In section \ref{33}, we review the derivation of GS light-cone 
superstring action from 
IIB matrix model, which is carried out in the previous paper \cite{KN3}.
The derivation of supersymmetry transformation for GS light-cone 
superstring from IIB matrix model is also shown.
The main investigation is carried out in 
section \ref{5} where
the vertex operators of IIB matrix model on the two
dimensional NC backgrounds are analyzed.
We verify that these operators are equivalent to 
superstring vertex operators.
Section \ref{6} is devoted to the conclusion.
Light-cone open superstring vertex operators are shown in 
appendix A.
In the appendix B, 
we construct type IIA closed string states in a radial quantization procedure.
By this construction, we can directly calculate the scattering amplitude 
from IIB matrix model.

\section{Vertex operators and supersymmetry transformation \label{3}}
\setcounter{equation}{0}

\subsection{Vertex operators in type IIB matrix model
\label{31}}

The Wilson loops are the vertex operators in IIB matrix model \cite{FKKT}.
The gauge invariant observables in noncommutative gauge theory are
the Wilson lines \cite{IIKK,Gross,DK} which are obtained from the 
Wilson loops in 
matrix models.
The behavior of closed string modes can be read from 
the correlation functions between the Wilson lines.
The various properties of NC gauge theory are investigated so far,
especially on homogeneous spaces \cite{homogeneous,KTT,KTT2}.
In four dimensional backgrounds,
the behavior of graviton propagators is investigated in detail 
through the vertex operators \cite{KN1}.

The vertex operators for the supergravity multiplet are constructed in
\cite{KMS}. 
These operators are linearly coupled to the supergravity fields and 
related with each other through the supersymmetry transformation.
The result which has been known up to now is summarized as follows.
$A_\mu \ (\mu=0,1,\cdots,9)$ and $\psi$ are $N \times N$ Hermitian
matrices and $\psi$ is a ten dimensional Majorana-Weyl spinor.

$\bullet$ Vertex operator for dilaton $\Phi$:
\begin{eqnarray} \label{vodilaton}
V^{\Phi}=\Str e^{ik\cdot A} \ ,
\end{eqnarray}
where the symmetric trace $\Str$ is defined as
\begin{eqnarray} \label{str}
\Str {\cal O}_1 {\cal O}_2 e^{ikA} \equiv
\int_0^1 d \alpha \tr {\cal O}_1 e^{i\alpha k A} {\cal O}_2
e^{i(1-\alpha) k A} \ .
\end{eqnarray}

$\bullet$ Vertex operator for dilatino $\tilde{\Phi}$:
\begin{eqnarray} \label{vodilatino}
V^{\tilde{\Phi}}= \Str e^{ik \cdot A}
 \bar{\psi} \ .
\end{eqnarray}

$\bullet$ Vertex 
operator for the second rank antisymmetric tensor field $B_{\mu \nu}$:
\begin{eqnarray} \label{voantisym2}
V_{\mu\nu}^B = \Str e^{ik\cdot A} \left(\frac{1}{16} \bar{\psi} \cdot \ks
\Gamma_{\mu \nu} \psi -\frac{i}{2} F_{\mu\nu} \right) \ ,
\end{eqnarray}
where 
\begin{eqnarray} 
F_{\mu\nu} \equiv [A_\mu,A_\nu] \ .
\end{eqnarray}

$\bullet$ Vertex operator for gravitino $\Psi_\mu$:
\begin{eqnarray} \label{vogravitino}
V_{\mu}^{\Psi}=
\Str e^{ik \cdot A} \left(
-\frac{i}{12} (\bar{\psi} \cdot \ks \Gamma_{\mu \nu} \psi)-2 F_{\mu\nu}\right)
\cdot \bar{\psi} \Gamma^{\nu} \ . \n
\end{eqnarray}

$\bullet$ Vertex operator for graviton $h_{\mu\nu}$:
\begin{eqnarray} 
V_{\mu\nu}^h =\Str e^{ik \cdot A} 
\left(-\frac{1}{96} k^\rho k^\sigma (\bar{\psi} \cdot \Gamma_{\mu
 \rho}^{\ \ \beta} \psi) \cdot (\bar{\psi} \cdot \Gamma_{\nu \sigma
 \beta} \psi)
\right.
\n 
-\frac{i}{4} k^\rho \bar{\psi} \cdot \Gamma_{\rho \beta (\mu } \psi
\cdot F_{\nu)}^{\ \beta}+\frac{1}{2} \bar{\psi} \cdot \Gamma_{(\mu}
[A_{\nu)},\psi]
\n
\left.
+2 F_{\mu}^{\ \rho} \cdot F_{\nu \rho}
\right) \ . \hspace*{1cm} \label{vograviton}
\end{eqnarray}

$\bullet$ Vertex operator for the fourth rank antisymmetric tensor
$C_{\mu\nu\rho\sigma}$:
\begin{eqnarray}
V_{\mu\nu\rho\sigma}^C=
\Str e^{ik\cdot A} \left(
\frac{i}{8 \cdot 4!} k_{\alpha} k_{\gamma}
(\bar{\psi} \cdot \Gamma_{[\mu\nu}^{\ \ \alpha} \psi)
\cdot (\bar{\psi} \cdot \Gamma_{\rho \sigma]}^{\ \ \gamma} \psi) 
\right. 
\n
+\frac{i}{3}\bar{\psi} \cdot \Gamma_{[\nu\rho \sigma} [\psi,A_{\mu]}]
+\frac{1}{4} F_{[\mu\nu} \cdot (\bar{\psi} \cdot \Gamma_{\rho\sigma]}^{\
\ \gamma} \psi)k_{\gamma}
\n
\left.
-i F_{[\mu\nu} \cdot F_{\rho \sigma]}
\right) \ . \hspace*{1cm} \label{voantisym4}
\end{eqnarray}

$\bullet$ Vertex operator for the conjugate gravitino $\Psi_{\mu}^c$:
\begin{eqnarray}
&&V^{\Psi^{c}}_{\mu}=
\n
     &&\Str e^{ik\cdot A}\bigg( -\frac{i}{2\cdot 5 !}
k^{\lambda}k^{\tau}(\bar{\psi}\cdot{\Gamma_{\mu \lambda}}^{\sigma}\psi)\cdot
(\bar{\psi}\cdot \Gamma_{\nu\tau\sigma}\psi)\cdot 
\bar{\psi}\Gamma^{\nu}\n
 &&+\frac{1}{24}k^{\lambda}(\bar{\psi}\cdot 
\Gamma_{\lambda\mu\nu}\psi)\cdot 
\bar{\psi}\Gamma^{\nu}\Gamma_{\rho\sigma}\cdot F^{\rho\sigma}
\n
&&
-\frac{1}{6}k^{\lambda}(\bar{\psi}\cdot \Gamma_{\lambda\alpha\beta}\psi)\cdot 
\bar{\psi}\Gamma^{\beta}\cdot {F^{\alpha}}_{\mu}\n
&&+i\bar{\psi}\Gamma_{\mu}[A_{\nu},\psi]\bar{\psi}\Gamma^{\nu}
-iF_{\mu\nu}\cdot F_{\rho\sigma}\cdot \bar{\psi}\Gamma^{\nu}
\Gamma^{\rho\sigma}\bigg) \ .
\label{vogravitinoc} 
\end{eqnarray}

$\bullet$ Vertex operator for the conjugate 
antisymmetric tensor field
$B_{\mu\nu}^c$:
\begin{eqnarray}
&&V^{B^{c}}_{\mu\nu}= \Str e^{ik\cdot A}
\n
&&\bigg( -\frac{1}{8 \cdot 6!}k^{\lambda}k^{\tau}k^{\alpha}(\bar{\psi}
 \cdot {\Gamma_{\mu\lambda}}^{\sigma}\psi) \cdot 
(\bar{\psi}\cdot \Gamma_{\gamma\tau\sigma} \psi)\cdot 
(\bar{\psi}\cdot {\Gamma^{\gamma}}_{\alpha\nu} \psi) \n
&& +\frac{i}{64}(\bar{\psi}\cdot\ks \Gamma_{\mu\alpha}\psi)\cdot 
F^{\alpha\beta}(\bar{\psi}\cdot \ks\Gamma_{\beta\nu}\psi)
\n
&&+\frac{i}{16\cdot 4!}(\bar{\psi}\cdot\ks \Gamma_{[\mu\alpha}\psi)
\cdot(\bar{\psi}\cdot\ks \Gamma^{\alpha\sigma}\psi)
\cdot        F_{\sigma\nu]}\n
&&-\frac{1}{32}\bar{\psi}\cdot\Gamma_{[\mu}[A^{\sigma},\psi]\cdot 
(\bar{\psi}\cdot\ks\Gamma_{\sigma\nu]}\psi) \n
&&-\frac{1}{64}(\bar{\psi}\cdot\ks\Gamma_{[\mu\alpha}\psi)\cdot 
\bar{\psi}\Gamma^{\alpha}[A_{\nu]},\psi] 
\n
&&+\frac{i}{4!\cdot 32}\Xi_{\mu\nu\alpha\beta\gamma}\cdot 
(\bar{\psi}\cdot \Gamma^{\alpha\beta\gamma}\psi) \n
&&-\frac{i}{64}[A_{\alpha},F^{\alpha\tau}]\cdot 
(\bar{\psi}\cdot \Gamma_{\tau\mu\nu}\psi) \n
&& +\frac{1}{64}(\bar{\psi}\cdot \Gamma_{\mu\nu\rho\sigma\lambda\tau}\ks 
\psi)\cdot F^{\rho\sigma}\cdot 
F^{\lambda\tau} \n
&&+\frac{1}{16}(\bar{\psi} \cdot \Gamma_{\rho\sigma}
\ks\psi)\cdot F^{\rho\sigma}\cdot F^{\mu\nu}\n
&&-\frac{1}{8}(\bar{\psi}\cdot \Gamma_{\rho\sigma}\ks \psi ) \cdot 
F^{\mu\rho}\cdot F^{\nu\sigma}
 +\frac{1}{8}(\bar{\psi}\cdot \Gamma_{[\mu\sigma}\ks \psi ) \cdot 
F^{\sigma\alpha}\cdot 
F_{\alpha\nu]}\n
&&-\frac{1}{32}(\bar{\psi}\cdot \Gamma^{\mu\nu}\ks \psi )\cdot 
F^{\rho\sigma}\cdot F_{\sigma\rho} 
+\frac{i}{4}\bar{\psi}\cdot \Gamma_{\mu\nu\alpha}[A_{\beta},\psi]\cdot 
F^{\alpha\beta}\n
 && +\frac{i}{8}\bar{\psi}\cdot \Gamma_{\rho\sigma[\mu}[A_{\nu]},\psi]
\cdot F^{\rho\sigma} 
+\frac{i}{4}\bar{\psi}\cdot \Gamma_{(\mu}[A_{\rho)},\psi] \cdot 
{F^{\rho}}_{\nu} \n
&&-\frac{i}{4}\bar{\psi}\cdot \Gamma_{(\nu}[A_{\rho)},\psi]\cdot
F^{\rho\mu} 
\n
&&-iF_{\mu\rho}\cdot F^{\rho\sigma}\cdot F_{\sigma\nu}
+\frac{i}{4}F_{\mu\nu}\cdot F^{\rho\sigma}\cdot F_{\sigma\rho}\bigg) \ ,
\label{voantisym2c}
\end{eqnarray}
where 
\begin{equation}
\Xi_{\mu\nu\rho\sigma\tau}=\{ \psi_{\alpha},\psi_{\beta} \} 
(\Gamma_0\Gamma_{\mu\nu\rho\sigma\tau})_{\alpha\beta} \ .
\end{equation}

The remaining vertex operators for the conjugate dilatino and dilaton
are only partly known.

$\bullet$ Vertex operator for the conjugate dilatino $\tilde{\Phi}^c$:
\begin{eqnarray}
&&V^{\tilde{\Phi}^c} = \Str e^{ik\cdot A} \n
&&\bigg(\frac{1}{8!}(\bar{\psi}\cdot\Gamma^{\alpha\gamma}\ks \psi)
\cdot (\bar{\psi}\cdot \Gamma_{\gamma\delta}\ks
\psi)\cdot(\bar{\psi}\cdot \Gamma^{\delta\beta}\ks \psi)\cdot
\bar{\psi} \Gamma_{\alpha\beta}\n
&&
 -\frac{i}{2\cdot 5!}F^{\mu\alpha}\cdot (\bar{\psi}\cdot 
\ks\Gamma_{\alpha\beta}\psi)
\cdot (\bar{\psi}\cdot \ks \Gamma^{\beta\nu}\psi)\cdot 
\bar{\psi}\Gamma_{\mu\nu} 
 \n
&&
+  \cdot \cdot \cdot -\frac{1}{8\cdot 4!}F^{\mu\nu}\cdot 
F^{\rho\sigma}(\bar{\psi}\cdot 
\Gamma_{\mu\nu\rho\sigma\lambda\alpha\beta}\psi)k^{\lambda}\cdot 
\bar{\psi}\Gamma^{\alpha\beta} \n
&&-\frac{1}{12}F^{\mu\alpha}\cdot F_{\alpha\beta}\cdot (\bar{\psi}\cdot 
\ks\Gamma^{\beta\nu}\psi)\cdot 
\bar{\psi}\Gamma_{\mu\nu} \n
&&
-\frac{1}{24}F^{\mu\alpha}\cdot (\bar{\psi}\cdot 
\ks\Gamma_{\alpha\beta}\psi)\cdot 
F^{\beta\nu}\cdot \bar{\psi}\Gamma_{\mu\nu} \n
&&-\frac{1}{48}
F^{\rho\sigma}\cdot (\bar{\psi}\cdot \ks 
\Gamma_{\rho\sigma}\psi)\cdot F^{\mu\nu}\cdot \bar{\psi}
\Gamma_{\mu\nu}+\cdot \cdot \cdot \n
&&
 +\cdot \cdot \cdot +\frac{i}{24}\bar{\psi}\cdot 
\Gamma_{\mu\nu\rho\sigma\lambda\tau}F^{\mu\nu}\cdot
F^{\rho\sigma}\cdot F^{\lambda\tau}\n
&&+i\bar{\psi}\cdot \Gamma^{\mu\nu}
(F_{\mu\rho}\cdot F^{\rho\sigma}\cdot F_{\sigma\nu}
-\frac{1}{4}F^{\rho\sigma}\cdot F_{\sigma\rho}\cdot F_{\mu\nu})
\bigg)\ . \n
\label{vodilatinoc}
\end{eqnarray}

$\bullet$ Vertex operator for the conjugate dilaton $\Phi^c$:
\begin{eqnarray}
&&V^{\Phi^c}=\Str e^{ik\cdot A} \n
&&\bigg(\frac{1}{8\cdot 8!}(\bar{\psi}\cdot \Gamma^{\alpha\gamma}\ks\psi)
\cdot (\bar{\psi} \cdot \Gamma_{\gamma\delta}\ks \psi)
\cdot (\bar{\psi}\cdot \Gamma_{\delta\beta}\ks\psi)\cdot 
(\bar{\psi}\cdot \Gamma_{\alpha\beta}\ks \psi) \nonumber \\
&&+\cdot \cdot \cdot \nonumber \\
&&  +\frac{i}{48}(\bar{\psi}\cdot 
\Gamma_{\mu\nu\rho\sigma\lambda\tau}\ks\psi ) 
\cdot F^{\mu\nu}\cdot F^{\rho\sigma}\cdot F^{\lambda\tau} \n
&& +[A_{\mu},\bar{\psi}]\cdot \Gamma_{\rho\sigma}\Gamma_{\nu}\psi 
\cdot F^{\mu\nu}\cdot F^{\rho\sigma}\nonumber \\
&&  +\frac{i}{2}(\bar{\psi}\cdot \Gamma^{\mu\nu}\ks \psi)\cdot 
(F_{\mu\rho}\cdot F^{\rho\sigma}\cdot F_{\sigma\nu}-\frac{1}{4}F^{\rho\sigma}
\cdot F_{\sigma\rho}\cdot F_{\mu\nu})  \nonumber \\
&&  -(F_{\mu\nu}\cdot F^{\nu\rho}\cdot F_{\rho\sigma}
\cdot F^{\sigma\mu}-\frac{1}{4}F_{\mu\nu}\cdot F^{\nu\mu}\cdot F_{\rho\sigma} 
\cdot F^{\sigma\rho})\bigg) \ . \n
\label{vodilatonc}
\end{eqnarray}

In these expressions, the Lorentz indices $\mu,\nu,\rho,\sigma,
\lambda,\tau,\alpha,\beta,\gamma,\delta$ run over $0,1,\cdots,9$.

\subsection{Vertex operators in superstring theory
\label{32}}

We review the construction of the vertex operators for Green-Schwarz
light-cone superstring.
In the early 1980s,
Green and Schwarz investigated the light-cone gauge formalism
of superstring theory.
The vertex operators are constructed in their formalism and the 
tree and one loop amplitudes are calculated \cite{GS1,GS2,GS3,GS4}. 

The light-cone coordinates $\phi_{i} (\tau,\sigma)$ decompose 
into the sum of the right-moving and left-moving components $\phi_{i}^R
(\tau-\sigma)+ \phi_{i}^L (\tau +\sigma)$.
Since the two sectors separately describe the Fock
space of open string states,
closed string states are given by the direct products
of open string states.
Open string vertex operators are shown in the appendix A.
There are $16 \times 16 = 256$ 
massless states in type IIA(IIB) superstring theory.
By the product of two SO(8) representations, or two super Yang-Mills
multiplets,
they are written as (left mover)$\times$(right mover)$=$ $(8_v + 8_c) \times (8_v+8_s)$ in type IIA superstring
and $(8_v + 8_s) \times (8_v +8_s)$ in type IIB superstring.
We focus on IIA superstring in what follows.

\underline{$8_v \times 8_v$ sector}

$8_v \times 8_v$ sector is decomposed into
\begin{eqnarray}
8_v \times 8_v =[0]+[2]+(2)=1+28+35_v \ ,
\end{eqnarray}
where [2] denotes the second rank antisymmetric tensor field $B_{ij}$
and (2) denotes the symmetric traceless tensor $h_{ij}$.
[0] corresponds to the dilaton $\Phi$.

The vertex operator for the symmetric traceless tensor in type IIA
superstring theory is given by \footnote{
In order to distinguish the superstring vertex operators from 
the vertex operators in IIB matrix model, we denote the closed superstring 
vertex operators in calligraphic characters.}
\begin{eqnarray}
({\cal V}_{ij})^h (k)&=& -\frac{1}{4 \pi \alpha'}
\int d \tau d \sigma 
V_{(i}^{B} 
(\frac{1}{2}k, \tau-\sigma)
 V_{j)}^{B} (\frac{1}{2}k,\tau +\sigma) \n 
&=& -\frac{1}{4 \pi \alpha'}\int d\tau d\sigma 
\left(\dot{ \phi}_R^{(i}  \dot{\phi}_L^{j)} 
-\frac{1}{8} \Gamma_{ab}^{(il} s_R^a s_R^b k^l \dot{\phi}_L^{j)}
\right.
\n
&&
-\frac{1}{8} \Gamma_{\dot{a}\dot{b}}^{(jl} s_L^{\dot{a}} s_L^{\dot{b}}
 k^l \dot{\phi}_R^{i)} 
 \n
&&\left. 
+\frac{1}{64} \Gamma_{ab}^{(il} s_R^a s_R^b k^l
\Gamma_{\dot{a} \dot{b} }^{j)m} s_L^{\dot{a}} s_L^{\dot{b}} k^m
\right) e^{i k \phi}  \ , \label{gra-str1}
\end{eqnarray}
where
\begin{eqnarray}
(\cdots )_R &\equiv& (\cdots )_R (\tau-\sigma) \ , \n
(\cdots )_L &\equiv& (\cdots )_L (\tau+\sigma) \ .
\end{eqnarray}
$V_i^B$ and $V_{1a}^F(V_{2\dot{a}}^F)$ 
denote the bosonic and fermionic 
vertex operators. The explicit form of these operators is 
shown in the appendix A.
The indices $i,j,l,m$ run over the transverse directions $2,\cdots,9$. 
The vertex operator for the second rank antisymmetric tensor is given by
\begin{eqnarray}
&&({\cal V}_{ij})^B (k)= \n
&& -\frac{1}{4 \pi \alpha'} \int d \tau d \sigma  V_{[i}^{B} 
(\frac{1}{2}k, \tau-\sigma)
 V_{j]}^{B} (\frac{1}{2}k,\tau +\sigma) 
 \ . \hspace*{1cm}
\label{b-str}
\end{eqnarray}
The vertex operator for dilaton is given by
\begin{eqnarray}
{\cal V}^\Phi (k)
&=& -\frac{1}{4 \pi \alpha'}  \int d \tau d \sigma V_{i}^{B} 
(\frac{1}{2}k, \tau-\sigma)
 V_{i}^{B} (\frac{1}{2}k,\tau +\sigma) \n 
&=& -\frac{1}{4 \pi \alpha'}\int d\tau d\sigma 
\left(\dot{ \phi}_R^{i}  \dot{\phi}_L^{i} 
-\frac{1}{8} \Gamma_{ab}^{il} s_R^a s_R^b k^l \dot{\phi}_L^{i}
\right.
\n
&&-\frac{1}{8} \Gamma_{\dot{a}\dot{b}}^{il} s_L^{\dot{a}} s_L^{\dot{b}}
 k^l \dot{\phi}_R^{i} 
 \n
&&\left. 
+\frac{1}{64} \Gamma_{ab}^{il} s_R^a s_R^b k^l
\Gamma_{\dot{a} \dot{b} }^{im} s_L^{\dot{a}} s_L^{\dot{b}} k^m
\right) e^{i k \phi} 
\ . \label{dil-str1}
\end{eqnarray}

\underline{$8_c \times 8_s$ sector}

The $8_c \times 8_s$ sector is decomposed into
\begin{eqnarray}
8_c \times 8_s =[1]+[3]=8_v+56 \ ,
\end{eqnarray}
where [1] and [3] denote the 1-form field $C_1^{i}$ and 3-form
antisymmetric tensor field $C_3^{ijl}$.

The vertex operator for the third rank antisymmetric tensor is given by
\begin{eqnarray}
&&({\cal V}_{ijl})^{C_3} (k)= -\frac{1}{4 \pi \alpha'}\int d \tau d
 \sigma 
\n&&\hspace*{0.5cm}
\left(  V_{1a}^{F} 
(\frac{1}{2}k, \tau-\sigma)_R \Gamma^{ijl}_{a\dot{b}} 
 V_{1\dot{b}}^{F} (\frac{1}{2}k,\tau +\sigma)_L \right)
\ , \hspace*{1cm}\label{3-form}
\end{eqnarray}
where $\Gamma^{ijl}_{a\dot{b}}$ 
is inserted to construct the irreducible tensor.
The vertex operator for the R-R 1-form field is given by
\begin{eqnarray}
&&({\cal V}_{i})^{C_1} (k)
= -\frac{1}{4 \pi \alpha'}\int d \tau d \sigma \n
&&\hspace*{0.5cm}\left(
 V_{1a}^{F} 
(\frac{1}{2}k, \tau-\sigma)_R \gamma^{i}_{a\dot{b}}
 V_{1\dot{b}}^{F} (\frac{1}{2}k,\tau +\sigma)_L \right) \ . 
\hspace*{1cm}
\label{1-form}
\end{eqnarray}

\underline{$8_v \times 8_s$ and $8_c \times 8_v$ sectors}

The representation $8_v \times 8_s$ ($8_c \times 8_v$) is 
decomposed into
\begin{eqnarray}
8_v \times 8_s &=&[1]+[3]=8_c +56_s \ , \n
8_c \times 8_v &=&[1]+[3]=8_s +56_c \ .
\end{eqnarray}
The vertex operator for gravitino is given by
\begin{eqnarray}
&&
({\cal V}_{i})^{\Psi} (k)=
 -\frac{1}{4 \pi \alpha'}
 \int d \tau d \sigma 
\n
&&\hspace*{1.5cm}
\left(
V_{1a}^{F} 
(\frac{1}{2}k, \tau-\sigma)_R 
 V_{i}^{B} (\frac{1}{2}k,\tau +\sigma)_L
 \right)
\ , \n
&&({\cal V}_{i})^{\Psi} (k)= 
-\frac{1}{4 \pi \alpha'}
 \int d \tau d \sigma 
\n
&&\hspace*{1.5cm}
\left(
V_{i}^{B} 
(\frac{1}{2}k, \tau-\sigma)_R 
 V_{1\dot{a}}^{F} (\frac{1}{2}k,\tau +\sigma)_L
 \right) \ . \hspace*{1cm}
\label{gravitino-str}
\end{eqnarray}

The vertex operator for dilatino is given by
\begin{eqnarray}
&&{\cal V}^{\tilde{\Phi}} (k)= -\frac{1}{4 \pi \alpha'}
\int d \tau d \sigma \n
&&\hspace*{0.8cm}\left( V_{1a}^{F} 
(\frac{1}{2}k, \tau-\sigma)_R \gamma^i_{a\dot{a}}
 V_{i}^{B} (\frac{1}{2}k,\tau +\sigma)_L \right) \ , \n
&&{\cal V}^{\tilde{\Phi}} (k)= -\frac{1}{4 \pi \alpha'}
\int d \tau d \sigma \n
&&\hspace*{0.8cm}
\left( V_{i}^{B} 
(\frac{1}{2}k, \tau-\sigma)_R \gamma^i_{a\dot{a}}
 V_{1\dot{a}}^{F} (\frac{1}{2}k,\tau +\sigma)_L \right) \ .
\hspace*{1cm}
\end{eqnarray}

\subsection{IIB matrix model and Green-Schwarz superstring \label{33}}

The action of IIB matrix model is written as
\begin{eqnarray}
S=-\frac{1}{g^2} \Tr
 \left(\frac{1}{4}[A^\mu,A^\nu][A_\mu,A_\nu]+\frac{1}{2}\bar{\psi} \Gamma^\mu
 [A_\mu,\psi] \right) \ . \n
\label{action1}
\end{eqnarray}
By expanding the matrices 
\begin{eqnarray}
A_\mu =p_\mu +a_\mu \ ,
\end{eqnarray}
around the two dimensional NC background,
\begin{eqnarray} \label{theta}
[p_\mu,p_\nu]=i\theta_{\mu\nu} \ ,
\end{eqnarray}
we obtain two dimensional noncommutative gauge theory with
${\cal N}=8$ supersymmetry \cite{CDS,AIIKKT,Li}
\begin{eqnarray}
&&S= \n
&&-\frac{\theta}{8 \pi g^2} \int d^2 x  \tr \left(
[D^{\tilde{\mu}},D^{\tilde{\nu}}][D_{\tilde{\mu}},D_{\tilde{\nu}}] +2
[D^{\tilde{\mu}},\phi^i][D_{\tilde{\mu}},\phi_i]
\right. \n
&&\left. +[\phi_i,\phi_j]
[\phi_i,\phi_j]
 +2 \bar{\psi} \Gamma^{\tilde{\mu}} [D_{\tilde{\mu}},\psi]
+2 \bar{\psi} \Gamma_i 
[\phi_i,\psi]
\right)_* , 
\label{2dNCYM}
\end{eqnarray}
where $\tilde{\mu},\tilde{\nu}=0,1$ and $i,j=2,\cdots ,9$.
\footnote{The metric is Wick rotated into the Euclidean signature  in order to 
make contact with NC gauge theory.
}
Trace of the matrices is mapped into the integral of the functions as
\begin{eqnarray} 
\Tr \to \frac{\theta}{2 \pi} \tr \int d^2 x  \ ,
\end{eqnarray}
where $\tr$ is a trace over $U(n)$ gauge group.
Noncommutative parameter $\theta$ is an off-diagonal matrix element 
of the matrix $\theta \equiv \theta_{01}$.

In the IR limit, 
we can identify the perturbative string spectrum.
Firstly, 

i) $*$ product reduces to ordinary commutative product
since higher derivatives in the product can be neglected.
The action (\ref{2dNCYM}) becomes the commutative ${\cal N}=8$
$U(n)$ super Yang-Mills in this limit
\begin{eqnarray}
S&=&-\frac{\theta}{8 \pi g^2} \int d^2 x \tr \left(
F^2_{\tilde{\mu}\tilde{\nu}} +2 (D_{\tilde{\mu}} \phi_i)^2
+[\phi_i,\phi_j][\phi_i,\phi_j] 
\right. \n &&\left.
+2 \bar{\psi} \Gamma^{\tilde{\mu}}
D_{\tilde{\mu}} \psi +2 \bar{\psi} \Gamma_i [\phi_i,\psi] \right) \ .
\label{2dYM}
\end{eqnarray}
This action includes 8 matrix scalar fields $\phi_i$ 
and 16 matrix spinor fields 
$\psi= (s^a, s^{\dot{a}})$. These fields transform 
in $8_v$, $8_c$ and $8_s$ representations of SO(8) group.
The perturbative vacua of this action are represented by
the diagonal matrices $\phi_i =(\phi_{\text{diag}})_i$, which 
form the moduli space of this theory.

By assuming that all the eigenvalues of matrices do not coincide 
with each other at any points on the worldsheet,
all excitations 
of the off-diagonal modes become massive.
Then, 

ii) only diagonal elements are relevant
in the low energy limit
since massless excitations come from the diagonal elements.
The interaction terms $[\phi_i,\phi_j][\phi_i,\phi_j]$ 
and $2 \bar{\psi} \Gamma_i
[\phi_i,\psi]$ vanish since the diagonal components commute. 
The gauge fields on the worldsheet decouple from the other fields.
It has been found that the IR limit 
corresponds to the free string limit \cite{DVV}.
It is therefore consistent to modify the short distance 
structure of their construction
as we have introduced the noncommutativity.

We transform the worldsheet coordinates from $R^2$ to $R^1 \times S^1$ as 
\begin{eqnarray} \label{conformal}
z \equiv x_0+ix_1=e^{\tau+i\sigma}  \ .
\end{eqnarray}
By the rescaling
\begin{eqnarray}
\psi_R \to \frac{1}{\sqrt{z}} \psi_R \ , \quad
\psi_L \to \frac{1}{\sqrt{\bar{z}}} \psi_L \ ,
\end{eqnarray}
we obtain an action for a single string with the winding number $w$
as a string may wind $w$ times in the $\sigma$ direction
\begin{eqnarray}
&&S=-\frac{\theta}{4 \pi g^2}
\int_{0}^{\infty} d\tau \int_0^{2\pi w} d\sigma \n
&& \left((\p_{\tau} \phi_i)^2 +(\p_\sigma \phi_i)^2
+ \bar{\psi} (\Gamma^+ \p_+ + \Gamma^- \p_- ) \psi
\right) \ . \label{GSaction2}
\end{eqnarray}
Since the rank of the gauge group is related with the 
winding number of the strings as $n=\sum_i w_i$,
multiple strings are obtained in general.
GS superstring action with light-cone gauge is obtained by identifying 
$\frac{\theta}{4 \pi g^2} \equiv \frac{1}{4 \pi \alpha'}$.

This action (\ref{GSaction2}) is invariant under the supersymmetry
transformation with 32 supercharges of type IIA string theory
which originates from ${\cal N}=2$
supersymmetry transformation in IIB matrix model
as follows.
Supersymmetry transformation in IIB matrix model
is written as
\begin{eqnarray}
\delta^{(1)} \psi &=&\frac{i}{2} [A_\mu,A_\nu]\Gamma^{\mu\nu} 
\epsilon \ , \n
\delta^{(1)} A_\mu &=&i \bar{\epsilon} \Gamma_\mu \psi \ , \n
\delta^{(2)}\psi &=&-\eta \ , \n
\delta^{(2)} A_\mu &=&0 \ .
\end{eqnarray}
On the two dimensional background,
this transformation reduces in the IR limit to 
\begin{eqnarray}
\delta^{(1)} s_a &=&- \dot{\phi}^i \gamma_{a\dot{a}}^i
\epsilon^{\dot{a}} \ , \quad
\delta^{(1)} s_{\dot{a}} =
- \dot{\tilde{\phi}}^i \gamma_{\dot{a}a}^i
\epsilon^{a}
\ ,
\n
\delta^{(1)} \phi_i &=&2
(\bar{\epsilon}^{\dot{a}} \gamma_{a\dot{a}}^i s^a+
\bar{\epsilon}^{a} \gamma_{\dot{a}a}^i s^{\dot{a}})
\ , \n
\delta^{(2)}s_a &=&- \eta^{a} \ , 
\quad
\delta^{(2)}s_{\dot{a}} =- \eta^{\dot{a}} \ , 
\n
\delta^{(2)} \phi_i &=&0 \ ,
\end{eqnarray}
where we have redefined $\eta^a \to \eta^a+\theta \epsilon$,
$\eta^{\dot{a}}\to \eta^{\dot{a}}-\theta \epsilon$ to 
absorb the constant shift.
The factors $\sqrt{z}$ and $\sqrt{\bar{z}}$ are absorbed by the
redefinition of $\epsilon$ and $\eta$.
$\gamma^i_{a\dot{a}}$ are the Clebsch-Gordan coefficients 
among three inequivalent SO(8) representations.
This transformation leaves the Green-Schwarz light-cone 
string action (\ref{GSaction2}) invariant.

\section{Superstring vertex operators in type IIB matrix model\label{5}}

\setcounter{equation}{0}

In this section,
we derive superstring vertex operators from those of IIB matrix model
on two dimensional backgrounds in the IR limit.
We verify the equivalence between our construction 
and the light-cone formalism (except for the dilaton and dilatino)
in subsections 3.1,3.2 and 3.3. 
For the dilaton and dilatino operators, we can find a
convincing correspondence between the both constructions
in section 3.4.

\subsection{$8_v \times 8_v$ sector}

\underline{Graviton $h_{ij}$}

The vertex operators for graviton in IIB matrix model is written as
\begin{eqnarray} 
V_{\mu\nu}^h =\Str e^{ik \cdot A} 
\left(-\frac{1}{96} k^\rho k^\sigma (\bar{\psi} \cdot \Gamma_{\mu
 \rho}^{\ \ \beta} \psi) \cdot (\bar{\psi} \cdot \Gamma_{\nu \sigma
 \beta} \psi)
 \right. \n 
-\frac{i}{4} k^\rho \bar{\psi} \cdot \Gamma_{\rho \beta (\mu } \psi
\cdot F_{\nu)}^{\ \beta}+\frac{1}{2} \bar{\psi} \cdot \Gamma_{(\mu}
[A_{\nu)},\psi]
\n
\left.
+2 F_{\mu}^{\ \rho} \cdot F_{\nu \rho}
\right) \ . \hspace*{1cm}
\end{eqnarray}
In the two dimensional background (\ref{theta}),
the vertex operators are written in terms of the fields 
$\phi^i$, $s^a$ and $s^{\dot{a}}$.
Furthermore, only the diagonal components are relevant in the 
IR limit.
The symmetric trace $\Str$ is mapped into the integral as
\begin{eqnarray}
\Str 
\rightarrow
\frac{\theta}{2\pi} \int_0^\infty d \tau \int_0^{2\pi w} d \sigma |z|^2 
(\cdots ) \ .
\end{eqnarray}
By the field redefinition
\begin{eqnarray}
s_a \to \frac{1}{\sqrt{wz}} s_a \ , \quad s_{\dot{a}} \to
\frac{1}{\sqrt{w\bar{z}}} s_{\dot{a}} \ , 
\label{redef}
\end{eqnarray}
and the scaling
\begin{eqnarray}
\tau \to w \tau \ , \quad \sigma \to w \sigma \ ,
\label{scaling}
\end{eqnarray}
we obtain the correspondence
\begin{eqnarray}
\Str \to \frac{\theta}{2\pi} \int_0^\infty d\tau \int_0^{2\pi} d\sigma
(\cdots ) \ .
\end{eqnarray}
In order to confirm that matrix model vertex operators are 
equivalent to superstring vertex operators,
we investigate the IR limit of the graviton vertex operators term by term.

(a) $F_{i}^{\ \mu} \cdot F_{j \mu}$ term

The bosonic part of the graviton vertex operator is written as
\begin{eqnarray} \label{gra-mat1}
2 \Str e^{ik \cdot A} F_{i}^{\ \mu} \cdot F_{j\mu} \ ,
\end{eqnarray}
where we can assume that the graviton has a transverse polarization.
In the two dimensional background,
leading contribution in the low energy limit
gives
\begin{eqnarray}
\Str e^{ik \cdot A} F_{(i}^{\ \mu} \cdot F_{j)\mu}
\to  \frac{\theta}{\pi} \int d^2 x e^{ik \cdot \phi} \p_- \phi_R^{(i}
\p_+ \phi_L^{j)} \ ,
\end{eqnarray}
where
\begin{eqnarray}
\p_{\pm}\equiv \p_\tau \pm i \p_\sigma  \ .
\end{eqnarray}
Thus, the operator (\ref{gra-mat1}) reduces to
the first term in (\ref{gra-str1}).

(b) $-\frac{i}{4} k^\rho \bar{\psi} \Gamma_{\rho \beta (i} \psi F_{j)}^{\
 \beta}$ term

Since $\Gamma_+$ and $\Gamma_-$ act on the fermion as
\begin{eqnarray}
\label{rel}
\bar{\psi}_R \Gamma_{+ij} \psi_R &=& i \Gamma_{ab}^{ij} s_R^a s_R^b \ , \n
\bar{\psi}_R \Gamma_{-ij} \psi_R &=& 0
\ , \n
\bar{\psi}_L \Gamma_{+ij} \psi_L &=& 0
\ , \n
\bar{\psi}_L \Gamma_{-ij} \psi_L &=& i \Gamma_{\dot{a}\dot{b}}^{ij} 
s_L^{\dot{a}} s_L^{\dot{b}} \ ,
\end{eqnarray}
we obtain 
\begin{eqnarray}
&&-\frac{i}{4}
\Str e^{ikA} k^\rho \bar{\psi} 
\Gamma_{\rho \beta (i} \psi F_{j)}^\beta
\n
&&= \frac{i}{8\pi}
\theta \int d\tau d \sigma e^{ik\phi} \left(k^l \bar{\psi} \Gamma_{l-(i} \psi
\dot{\phi}_R^{j)}+k^l \bar{\psi} \Gamma_{l+(i} \psi
\dot{\phi}_L^{j)} \right)
 \n 
&&= -\frac{\theta}{8\pi}
\int d\tau d\sigma e^{i k \phi}  
\left( 
\Gamma_{ab}^{(il} s_R^a s_R^b k^l \p_+ {\phi}_L^{j)} 
\right. \n
&&\hspace*{3.5cm}\left.
+ \Gamma_{\dot{a}\dot{b}}^{(il} s_L^{\dot{a}} s_L^{\dot{b}} 
k^l \p_- {\phi}_R^{j)} 
\right) \ , 
\end{eqnarray}
by the field redefinition (\ref{redef}).
Thus, this term is equivalent to the
 second and third terms in (\ref{gra-str1}).

(c) $-\frac{1}{96}
k^\rho k^\sigma ( \bar{\psi} \Gamma_{i \rho}^{\ \ \beta} \psi )
( \bar{\psi} \Gamma_{j \sigma \beta}\psi)$ term

This term reduces to
\begin{eqnarray}
&&-\frac{1}{192 }
\Str e^{ikA}k^\rho k^\sigma ( \bar{\psi} \Gamma_{(i \rho}^{\ \ \beta} \psi )
( \bar{\psi} \Gamma_{j)\sigma \beta}\psi) \n
&=&
-\frac{\theta}{384\pi} \int d\tau d \sigma
 e^{ik\phi} \left(
k^l k^m ( \bar{\psi} \Gamma_{(i l}^{\ \ +} \psi )
( \bar{\psi} \Gamma_{j)m}^{\ \ \ -} \psi) 
\right. \n 
&& \left.
+k^l k^m ( \bar{\psi} \Gamma_{(i l}^{\ \ -} \psi )
( \bar{\psi} \Gamma_{j)m}^{\ \ \ +} \psi) 
\right. \n
&& \left. +
k^l k^m ( \bar{\psi} \Gamma_{(i l}^{\ \ n} \psi )
( \bar{\psi} \Gamma_{j)m n}\psi) 
\right. \n
&&\left. +
 k^- k^- ( \bar{\psi} \Gamma_{(i -}^{\ \ \ n} \psi )
( \bar{\psi} \Gamma_{j)- n}\psi) \right) \n
&=& \frac{\theta}{384\pi} \int d\tau d \sigma e^{ik \phi}
\left(2 \Gamma_{ab}^{(il} s_R^a s_R^b k^l
\Gamma_{\dot{a}\dot{b} }^{j)m} s_L^{\dot{a}} s_L^{\dot{b}} k^m \right. \n
&&
+4 \Gamma_{a\dot{a}}^{(iln} s_R^a s_L^{\dot{a}} k^l
\Gamma_{b\dot{b}}^{j)mn} s_R^b s_L^{\dot{b}} k^m
\n &&\left.
+ k^- k^- \Gamma_{ab}^{in} s_a s_b \Gamma_{cd}^{mn}
s_c s_d
\right) \n
&=& \frac{\theta}{64\pi} \int d\tau d\sigma e^{ik \phi}
\Gamma_{ab}^{(il} s_R^a s_R^b k^l
\Gamma_{\dot{a}\dot{b} }^{j)m} s_L^{\dot{a}} s_L^{\dot{b}} k^m 
\ , 
\label{gra-mat3}
\end{eqnarray}
which is equivalent to the last term in (\ref{gra-str1}).

(d) $\frac{1}{2} \bar{\psi} \cdot \Gamma_{(i}
[A_{j)},\psi]$ term

This term vanishes in the IR limit.

\

Next, let us consider the light cone momentum in the exponential 
factor $k \cdot \phi =k_i \phi_i -k^- \phi^+ -k^+ \phi^-$.\footnote{
In the light-cone gauge evaluations, we put $k^+=0$
except for the initial and final states.
}
Gauge fields exist in this factor as
\begin{eqnarray}
k^- \phi^+ +k^+ \phi^- =k^- (p^+ +a^+) +k^+ (p^- +a^-) \ . \n
\end{eqnarray}
The correlation function between the gauge fields is given by
\begin{eqnarray}
\langle e^{i k_1^- a_1^+ }(z_1) \cdot e^{i k_2^+ a_2^- )}(z_2) \rangle \sim
|z_1-z_2|^{-\alpha'  k_1^-  k_2^+ } \ , \n
\label{ope}
\end{eqnarray}
where $\alpha'=g^2/\theta$.
If we put together the correlation functions between 
the gauge fields and scalar fields,
the momentum dependent power of (\ref{ope}) is summed up in the covariant 
form.
It is because we started with the covariant IIB matrix model action.

On the other hand, in the light-cone gauge formalism, 
$k \cdot \phi=k_i \phi_i -k^- \phi^+ =k_i
\phi_i -k^- \alpha' (p^+ \tau +x^+) $ where $k^+=0$.
After rotating $\tau\rightarrow -i\tau$, 
the light-cone momentum in the vertex operator
gives a factor 
\begin{eqnarray}
e^{-\alpha' k_1^-  p^+ \tau_1 
} | k_2 \rangle
&\sim& e^{-\alpha' \tau_1 k_1^-  k_2^+} |k_2 \rangle \n
&\sim& |z_1-z_2|^{ -\alpha' k_1^-  k_2^+} | k_2 
\rangle 
\end{eqnarray}
to the scattering amplitude.
This factor is also summed up in the covariant form 
if we put together the contribution from the transverse and longitudinal 
modes.
Thus, the light-cone momentum contributes consistently
to the amplitude
in the both formalism.

In this way, through the calculations (a),(b),(c) and (d), 
we have verified that the graviton vertex operator in IIB matrix model
(\ref{vograviton}) reduces in the IR limit to
\begin{eqnarray}
V_{ij}^h&=& 
\Str e^{ik \cdot A} 
\left(-\frac{1}{96} k^\rho k^\sigma (\bar{\psi} 
\cdot \Gamma_{i
 \rho}^{\ \ \beta} \psi) \cdot (\bar{\psi} 
\cdot \Gamma_{j \sigma
 \beta} \psi)
 \right. \n 
&&
-\frac{i}{4} k^\rho \bar{\psi} \cdot \Gamma_{\rho \beta 
(i } \psi
\cdot F_{j)}^{\ \beta}+\frac{1}{2} \bar{\psi} 
\cdot \Gamma_{(i}
[A_{j)},\psi]
\n &&\left.
+2 F_{i}^{\ \rho} \cdot F_{j \rho}
\right) \n
&\to&\frac{\theta}{\pi} \int_0^\infty d \tau \int_0^{2\pi} 
d \sigma e^{ik \cdot\phi}
\left(\dot{ \phi}_R^{(i}  \dot{\phi}_L^{j)} 
-\frac{1}{8} \Gamma_{ab}^{(il} s_R^a s_R^b k^l \dot{\phi}_L^{j)}
\right.
\n&&
-\frac{1}{8} \Gamma_{\dot{a}\dot{b}}^{(jl} s_L^{\dot{a}} s_L^{\dot{b}}
 k^l \dot{\phi}_R^{i)}  \n &&\left.
+\frac{1}{64} \Gamma_{ab}^{(il} s_R^a s_R^b k^l
\Gamma_{\dot{a} \dot{b} }^{j)m} s_L^{\dot{a}} s_L^{\dot{b}} k^m
\right)
\n
&=& ({\cal V}_{ij})^h
 \ ,  \label{gvo}
\end{eqnarray}
which is the graviton vertex operator
in type IIA superstring theory.

\underline{Second rank antisymmetric tensor $B_{ij}$}

The vertex operator for $B_{ij}$ in type IIB matrix model is shown in
(\ref{voantisym2}) and (\ref{voantisym2c}).
We will verify that (\ref{voantisym2c}) in the two dimensional background
is equivalent to (\ref{b-str}), which is 
$B_{ij}$ type superstring vertex operator.
For the graviton vertex operators, we have explicitly checked the
equivalence term by term. But
since (\ref{voantisym2c}) consists of many terms
compared to (\ref{vograviton}), we verify the
equality in a systematic way. 

The operator in IIB matrix model which reduces to $B_{ij}$
type (1,1) operator should be either of the following four forms
\begin{eqnarray}
&&(\bar{\psi} \Gamma_{i \cdots} \psi)_L (\bar{\psi} \Gamma_{j \cdots} \psi)_R
 \cdots \ , \n
&&(\bar{\psi} \Gamma_{i \cdots} \psi)_L 
(F_{j\cdots})_R \cdots \ , \n
&&(F_{i\cdots})_L (\bar{\psi} \Gamma_{j \cdots} \psi)_R 
 \cdots \ , \n
&&(F_{i \cdots})_L (F_{j \cdots})_R  \cdots \ . \label{four-types}
\end{eqnarray}
They should not contain no extra
dimensionful operators. 
The terms in (\ref{voantisym2c}) 
which are of the forms (\ref{four-types}) are only 
the following three terms
\begin{eqnarray}
&& \Str e^{ik\cdot A}
\left( 
 \frac{i}{64}(\bar{\psi}\cdot\ks \Gamma_{\mu\alpha}\psi)\cdot 
F^{\alpha\beta}(\bar{\psi}\cdot \ks\Gamma_{\beta\nu}\psi)
\right.
\n
&&
+\frac{1}{8} (\bar{\psi} \cdot \Gamma_{\mu\sigma} \ks \psi) \cdot
F^{\sigma \alpha} \cdot F_{\alpha \nu} \n
&&\left.
- i F_{\mu\rho}\cdot F^{\rho\sigma}\cdot F_{\sigma\nu} 
\right) \ . \label{2-form}
\end{eqnarray} 
Other terms do not reduce to the (1,1) operators and do not give 
any contributions to the amplitude in the IR limit.
By the field redefinition (\ref{redef}) and the scaling (\ref{scaling}),
we find that
these terms are equivalent to 
\begin{eqnarray}
&&\frac{\theta}{\pi} 
\int d\tau d \sigma e^{ik \phi}
\left( 
\p_- { \phi}_R^{[i}  \p_+ {\phi}_L^{j]} 
-\frac{1}{8} \Gamma_{ab}^{[il} s_R^a s_R^b k^l \p_+ {\phi}_L^{j]}
\right.
\n
&&-\frac{1}{8}  \Gamma_{\dot{a}\dot{b}}^{[il} s_L^{\dot{a}} \p_- {\phi}_R^{j]}
s_L^{\dot{b}} k^l 
 \n
&&\left.+\frac{1}{64} \Gamma_{ab}^{[il} s_R^a s_R^b k^l
\Gamma_{\dot{a} \dot{b} }^{j]m} s_L^{\dot{a}} s_L^{\dot{b}} k^m
\right) \ ,
\end{eqnarray}
which is indeed the $B_{ij}$ superstring vertex operator (\ref{b-str}).
Thus, 
we have verified that, in the two dimensional background,
the antisymmetric two form vertex operator in IIB matrix model 
(\ref{voantisym2c}) reduces to
\begin{eqnarray}
\frac{i}{\theta}V_{ij}^{B^c} &=& \frac{\theta}{\pi} 
\int d\tau d \sigma e^{ik \phi}
\left( 
\p_- { \phi}_R^{[i}  \p_+ {\phi}_L^{j]} 
\right. \n
&&
-\frac{1}{8} \Gamma_{ab}^{[il} s_R^a s_R^b k^l \p_+ {\phi}_L^{j]}
-\frac{1}{8}  \Gamma_{\dot{a}\dot{b}}^{[il} s_L^{\dot{a}} \p_- {\phi}_R^{j]}
s_L^{\dot{b}} k^l 
 \n
&&\left.+\frac{1}{64} \Gamma_{ab}^{[il} s_R^a s_R^b k^l
\Gamma_{\dot{a} \dot{b} }^{j]m} s_L^{\dot{a}} s_L^{\dot{b}} k^m
\right) \ . \label{result}
\end{eqnarray}

\underline{Dilaton $\Phi$}

Although the dilaton vertex operators in IIB matrix model are not completely
determined yet, we can find a convincing correspondence between the
matrix model
vertex operators and the superstring vertex operators. We will discuss 
this consistency in section 3.4.

\subsection{$8_c \times 8_s$ sector}

Since we start from the type IIB supergravity multiplet, the R-R sector
includes 2-form and 4-form fields.
From them, we can obtain 1-form and 3-form fields 
by singling out the light-cone (-) direction.

\underline{The third rank antisymmetric tensor $C_{ijl}$}

Vertex operator for the 4-form field in IIB matrix model 
is present in (\ref{voantisym4}) as
\begin{eqnarray}
V_{\mu\nu\rho\sigma}^C=
\Str e^{ik\cdot A} \left(
\frac{i}{8 \cdot 4!} k_{\alpha} k_{\gamma}
(\bar{\psi} \cdot \Gamma_{[\mu\nu}^{\ \ \alpha} \psi)
\cdot (\bar{\psi} \cdot \Gamma_{\rho \sigma]}^{\ \ \gamma} \psi)
\right.
\n
+\frac{i}{3}\bar{\psi} \cdot \Gamma_{[\nu\rho \sigma} [\psi,A_{\mu]}]
+\frac{1}{4} F_{[\mu\nu} \cdot (\bar{\psi} \cdot \Gamma_{\rho\sigma]}^{\
\ \gamma} \psi)k_{\gamma}
\n 
\left. 
-i F_{[\mu\nu} \cdot F_{\rho \sigma]}
\right) \ . \hspace*{1cm}\label{4to3}
\end{eqnarray}
We can obtain the operators for 4-form, 3-form and 2-form fields
from the vertex operator (\ref{4to3}) as
\begin{eqnarray}
(V_{ijlm})^{C} \ , \label{4form} \\
(V^-_{\ \ ijl})^{C} \ , \label{3form} \\
(V^+_{\ \ ijl})^{C} \ , \\
(V^{+-}_{\ \ \ ij})^{C} \ . \label{2form}
\end{eqnarray}
We will show that only the 3-form (\ref{3form}) survives in the IR
limit. 

First of all, we discuss the 4-form vertex operator
(\ref{4form}).
The first term in the operator (\ref{4form}) is decomposed into
\begin{eqnarray}
&&\Str e^{ik\cdot A} 
\frac{i}{8 \cdot 4!} \left(
k_{n} k_{p}
(\bar{\psi} \cdot \Gamma_{[ij}^{\ \ n} \psi)
\cdot (\bar{\psi} \cdot \Gamma_{lm]}^{\ \ p} \psi)\right. \n
&&+k_{+} k_{p}
(\bar{\psi} \cdot \Gamma_{[ij}^{\ \ +} \psi)
\cdot (\bar{\psi} \cdot \Gamma_{lm]}^{\ \ p} \psi) \n
&&+k_{-} k_{p}
(\bar{\psi} \cdot \Gamma_{[ij}^{\ \ -} \psi)
\cdot (\bar{\psi} \cdot \Gamma_{lm]}^{\ \ p} \psi) \n
&&+k_{n} k_{+}
(\bar{\psi} \cdot \Gamma_{[ij}^{\ \ n} \psi)
\cdot (\bar{\psi} \cdot \Gamma_{lm]}^{\ \ +} \psi) \n
&&+k_{n} k_{-}
(\bar{\psi} \cdot \Gamma_{[ij}^{\ \ n} \psi)
\cdot (\bar{\psi} \cdot \Gamma_{lm]}^{\ \ -} \psi) \n
&&\left.
+2 k_{+} k_{-}
(\bar{\psi} \cdot \Gamma_{[ij}^{\ \ +} \psi)
\cdot (\bar{\psi} \cdot \Gamma_{lm]}^{\ \ -} \psi)
\right) \ . \label{4-1}
\end{eqnarray}
The first term and the last term in (\ref{4-1}) 
vanish because $i,j,l$ and $m$
are antisymmetric. Other terms do not contain (1,1) operators in 
the IR limit. Thus, this term does not contribute 
to the amplitude.
The second term in the 4-form (\ref{4form}) vanish in the IR limit
since $\psi$ and $A_i$ commute.
The third and fourth terms in (\ref{4form}) apparently do not contain (1,1) 
operator in the IR limit.
Thus, we have confirmed that 4-form (\ref{4form})
does not contribute to the 
amplitude.

In order to obtain the 3-form field, 
we put $\mu$ direction with the $-$ direction.
Then, the first, third and fourth terms in (\ref{3form}) 
vanish from the condition
that the indices $i,j$ and $l$ are antisymmetrized.
By redefining the field as (\ref{redef}) and (\ref{scaling}),
the third term in 
(\ref{3form}) reduces to
\begin{eqnarray}
(V_{\ \ ijl}^{-})^{C_4}&=&\Str e^{ik \cdot \phi} \frac{i}{3} 
\bar{\psi} \cdot \Gamma_{[ijl]} [ \psi,A^-] \n
&=&\frac{i\theta}{3\pi} 
\int d\tau d\sigma e^{ik\phi} s_R^a \Gamma_{a\dot{b}}^{[ijl]}
s_L^{\dot{b}} |wz| [k \phi ,p^-]_* \n 
&=&
k^- \frac{\theta }{3\pi} \int d\tau d \sigma s_R^a \Gamma_{a\dot{b}}^{[ijl]}
s_L^{\dot{b}} e^{ik\phi} \frac{|p^+|}{\theta} \ . 
\end{eqnarray}
In the last line, we use the canonical commutation relation $[p^+,p^-]=\theta$.
The initial state can be represented by a coherent state $\exp (k^+
\hat{p}^- ) | 0 \rangle$. Since $w \hat{z}^+ =\hat{p}^+ /
\theta$ on a NC plane, $w \hat{z}^+$ is fixed to be $w \hat{z}^+ 
=|p^+|/\theta$. Semiclassically we also find that
$w\hat{z}^-  = |p^+|/\theta $ since $|p^+|$ is real.
In this way, we have obtained the 3-form field in $8_c \times 8_s$ sector.
Thus, we have verified that
the vertex operator
(\ref{3form}) reduces to the superstring vertex operator (\ref{3-form}).
In this reduction, fermions have dimension $(\frac{1}{2},\frac{1}{2})$ and
the background $A^- \to p^-$ is interpreted to have dimension 
$(\frac{1}{2},\frac{1}{2})$.

If we single out the opposite direction $+$, we obtain the 3-form
field $(V^+_{\ \ ijl})^C$. This operator vanishes
\begin{eqnarray}
(V_{\ \ ijl}^{+})^{C_4}&=&\Str e^{ik \cdot \phi} \frac{i}{3} 
\bar{\psi} \cdot \Gamma_{[ijl]} [ \psi,A^+]  \n
&=&\frac{i\theta}{3\pi} 
\int d\tau d\sigma e^{ik\phi} s_R^a \Gamma_{a\dot{b}}^{[ijl]}
s_L^{\dot{b}} |wz| [k \phi ,p^+]_* \n
&=&0 \ ,
\end{eqnarray}
since $k^+=0$.

Finally, by the properties that Lorentz indices are antisymmetric,
one can confirm that 2-form (\ref{2form})
does not contribute to the amplitude.

In this way, 
we have verified that
the 4-form matrix model vertex operator
(\ref{voantisym2c}) reduces to the 3-form 
superstring vertex operator (\ref{3-form})
by singling out the $-$ direction.

\underline{One form field $C_i$}

From (\ref{voantisym2c}), 
we can obtain the operators for 2-form, 1-form and scalar fields as
\begin{eqnarray}
(V_{ij})^B \ , \label{2-form2}\\
(V^-_{\ \ i})^B \label{1-form2}\ , \\
(V^+_{\ \ i})^B \label{1-form2+}\ , \\
(V^{+-})^B \ . \label{0-form2}
\end{eqnarray}
2-form (\ref{2-form2}) reduces to $B_{ij}$ superstring vertex operator 
(\ref{b-str}) as we have verified in the previous section. 

1-form (\ref{1-form2}) and (\ref{1-form2+}) reduce to
\begin{eqnarray}
\frac{i}{\theta} (V^{-}_{\ \ i})^B&=&
\frac{-1}{2\theta} \Str e^{i k \cdot A} \bar{\psi} \cdot \Gamma^{-}_{\ i\alpha}
[A_{\beta},\psi] \cdot F^{\alpha \beta} \n 
&=&
-\frac{i}{2\pi}\int d\tau d\sigma e^{ik\phi} s_R^a\gamma_{a\dot{b}}^i
s_L^{\dot{b}} |wz| [p^-,k \phi]_* \n
&=& 
k^- \frac{ \theta }{2\pi} \int d\tau d \sigma e^{ik\cdot \phi} s_R^a \gamma_{a
\dot{b}}^i s_L^{\dot{b}} \frac{|p^+|}{\theta} \ , \\
\frac{i}{\theta} (V^{+}_{\ \ i})^B&=&0 \ .
\end{eqnarray}
Other terms in (\ref{1-form2}) do not
contribute to the amplitude in the IR limit.
One can also confirm that 0-form (\ref{0-form2}) does not contribute 
to the amplitude.
The operators in (\ref{voantisym2}) do not contribute or they have 
the dimension $(\frac{1}{2},\frac{1}{2})$, $(1,0)$ or $(0,1)$.

Thus, we have verified that matrix model vertex operator (\ref{voantisym2c})
reduces to the superstring vertex operator (\ref{1-form}) by
singling out the $-$ direction.

\subsection{$8_v \times 8_s$ ($8_c \times 8_v$) sector}

\underline{Gravitino $\Psi_i$}

The vertex operator for gravitino $\Psi_{\mu}$ in IIB matrix model is shown in
(\ref{vogravitino}) and (\ref{vogravitinoc}).
The vertex operator (\ref{vogravitino}) is
\begin{eqnarray}
V_{\mu}^{\Psi}=\Str e^{ik \cdot A} \left(
-\frac{i}{12} (\bar{\psi} \cdot \ks \Gamma_{\mu \nu} \psi)-2 F_{\mu\nu}\right)
\cdot \bar{\psi} \Gamma^{\nu} \ . \n
\end{eqnarray} 
By the field redefinition (\ref{redef}) and the scaling 
(\ref{scaling}), it reduces to
\begin{eqnarray}
V_{i}^{\Psi}
&&=  \n
&&-\frac{\theta}{2\pi} \int d\tau d\sigma
 e^{ik\phi} \left( \left(
-2\dot{\phi}^{i} +\frac{1}{4} \Gamma_{ab}^{i l} s^a  s^b k^l
\right)_R s_L^{\dot{a}} \sqrt{wz} 
\right.
\n
&&
\left.
+\sqrt{w\bar{z}} s_R^a  \left(
-2\dot{\phi}^{i} +\frac{1}{4} \Gamma_{\dot{a}\dot{b}}^{i l} s^{\dot{a}}
  s^{\dot{b}} k^l
\right)_L \right) \n
&&=  \frac{\theta}{\pi} \int d\tau d \sigma
 e^{ik\phi} \left( \left(
\dot{\phi}^{i} -\frac{1}{8} \Gamma_{ab}^{i l} s^a  s^b k^l
\right)_R \sqrt{\frac{p^+}{ \theta}}
 s_L^{\dot{a}}
\right.
\n
&&
\left.
+\sqrt{\frac{p^+}{ \theta}}
s_R^a  \left(
\dot{\phi}^{i} -\frac{1}{8} \Gamma_{\dot{a}\dot{b}}^{i l} s^{\dot{a}}
  s^{\dot{b}} k^l 
\right)_L \right) \ . 
\end{eqnarray}
In this way, we have shown that (\ref{vogravitino}) is equivalent to
the superstring vertex operator in 
(\ref{gravitino-str}).
One can check that the matrix model vertex operators 
for the conjugate gravitino (\ref{vogravitinoc}) 
do not contribute to the amplitude in the IR limit.

\subsection{Dilaton and dilatino}

\underline{Dilaton $\Phi$}

The dilaton vertex operator in IIB matrix model is shown in
(\ref{vodilaton}) 
\begin{eqnarray} \label{dil-ver}
\Str e^{ik \cdot A} \ ,
\end{eqnarray}
and (\ref{vodilatonc}) 
\begin{eqnarray}
&&V^{\Phi^c}=\Str e^{ik\cdot A} \n
&&\left(
  \frac{i}{2}(\bar{\psi}\cdot \Gamma^{\mu\nu}\ks \psi)\cdot 
(F_{\mu\rho} \cdot F^{\rho\sigma} \cdot F_{\sigma\nu}-\frac{1}{4}
F^{\rho\sigma}
\cdot F_{\sigma\rho}\cdot F_{\mu\nu}) \right. \n
&&   -
\left(F_{\mu\nu}\cdot F^{\nu\rho} \cdot F_{\rho\sigma} \cdot F^{\sigma\mu}
\right. \n
&&\left. \left.
-\frac{1}{4}F_{\mu\nu}\cdot F^{\nu\mu}\cdot F_{\rho\sigma} 
\cdot F^{\sigma\rho}\right)+\cdots \right) \ .
\label{dil-ver-conj}
\end{eqnarray} 
In the two dimensional background, 
by the field redefinition (\ref{redef}) and the scaling (\ref{scaling}), 
the explicitly written terms in (\ref{dil-ver-conj}) reduce to
\begin{eqnarray}
\frac{\theta}{\pi}  
\int d\tau d\sigma e^{ik \phi}
\left( 
-\frac{1}{8} \Gamma_{ab}^{il} s_R^a s_R^b k^l \p_+ {\phi}_L^{i}
\right.
\n
\left.
-\frac{1}{8} \p_- {\phi}_R^{i} \Gamma_{\dot{a}\dot{b}}^{il}
s_L^{\dot{a}} s_L^{\dot{b}} k^l 
+  \p_- \phi^i_R \p_+ \phi^i_L 
\right) \ .
\end{eqnarray}
These terms are present in (\ref{dil-str1}) and the relative
numerical coefficients also agree.
But the last term in the superstring vertex operator 
(\ref{dil-str1}) is missing. This term may come
from the undetermined terms in (\ref{vodilatonc}), for example,
\begin{eqnarray}
\Str e^{ik \cdot A}(\bar{\psi} \cdot \Gamma_{\lambda \rho} \ks \psi )
(\bar{\psi} \cdot \Gamma^{\lambda \rho} \ks \psi )
\cdot F^{\mu\nu} F_{\nu \mu} \ . \label{ex}
\end{eqnarray}
We also need to examine the other terms, that is, 
$(\cdots)$ in (\ref{dil-ver-conj}).
Explicitly determined terms in the (conjugate) dilaton vertex operators 
in the paper \cite{KMS} is (\ref{dil-ver-conj}) and 
\begin{eqnarray}
&&\Str e^{ik\cdot A} \n
&&\left(
\frac{1}{8\cdot 8!}(\bar{\psi}\cdot \Gamma^{\alpha\gamma}\ks\psi)
\cdot (\bar{\psi} \cdot \Gamma_{\gamma\delta}\ks \psi)
\cdot (\bar{\psi}\cdot \Gamma_{\delta\beta}\ks\psi)\cdot 
(\bar{\psi}\cdot \Gamma_{\alpha\beta}\ks \psi) \right. \n
&&   +\frac{i}{48}(\bar{\psi}\cdot 
\Gamma_{\mu\nu\rho\sigma\lambda\tau}\ks\psi ) 
\cdot F^{\mu\nu}\cdot F^{\rho\sigma}\cdot F^{\lambda\tau} \n
&& \left.
 +[A_{\mu},\bar{\psi}]\cdot \Gamma_{\rho\sigma}\Gamma_{\nu}\psi 
\cdot F^{\mu\nu}\cdot F^{\rho\sigma} \right)
\ . \label{dil2}
\end{eqnarray}
Therefore, we should check the behavior of these terms.
Since the total dimension of the first term in (\ref{dil2})
is four, there are no (1,1) operators in this term. 
In order to obtain (1,1) operators from the second term in (\ref{dil2}), 
we need to fix $F^{\mu\nu} =F^{\rho\sigma} \equiv F^{+-}$. 
Apparently, this term does
not contribute to the amplitude since it vanishes by the
$\Gamma_{+-+-} $ projection. 
The last term in (\ref{dil2}) does not survive by the similar reason.
Unless at least two of $\rho,\sigma$ and $\nu$ are transverse
indices, we cannot find the (1,1) operator. But 
if we choose $\rho$ and $\sigma$ to be the transverse directions as
$\rho=i$ and $\sigma=j$,
total dimension of the operator becomes three.
Thus, we have confirmed that the other terms than (\ref{dil-ver-conj})
which are explicitly 
constructed in
\cite{KMS} do not contribute to the amplitude.
Although there are many unknown terms in the matrix model vertex operators 
(\ref{vodilatonc}), such operators should reduce to the (1,1) operators 
in (\ref{dil-str1}). We thus believe that all the terms
boil down to the superstring vertex operators 
(\ref{dil-str1}).

\underline{Dilatino $\tilde{\Phi}$}

The vertex operator for $\Psi_{\mu}$ in type IIB matrix model is shown in
(\ref{vodilatino}) and (\ref{vodilatinoc}).
The dilatino vertex operator in (\ref{vodilatinoc}) reduces to that
in superstring as
\begin{eqnarray}
&&-\frac{1}{\theta^2}
V^{\tilde{\Phi}^c} 
\n
&&=- \frac{1}{\theta^2} \Str e^{ik\cdot A} \left( 
 -  \frac{1}{12} F^{\mu\alpha} \cdot F_{\alpha\beta}
\cdot (\bar{\psi} \cdot \ks \Gamma^{\beta\nu} \psi) \cdot \bar{\psi}
\Gamma_{\mu\nu} \right. \n
&&\left. + i\bar{\psi}
\cdot \Gamma^{\mu\nu}
(F_{\mu\rho} \cdot F^{\rho\sigma} \cdot F_{\sigma \nu}-
\frac{1}{4}F^{\rho\sigma} \cdot F_{\sigma \rho} \cdot F_{\mu\nu}) 
\right) \n
&&=\frac{\theta}{\pi} \int d\tau d\sigma e^{ik\phi}
\sqrt{\frac{p^+}{ \theta}}
\left(
-\frac{1}{8} \Gamma^{ij}_{bc} s_R^b s_R^c k^j \Gamma_{a\dot{a}}^i 
s_L^{\dot{a}}
\right.
\n
&&\left.
+ \dot{\phi}_R \Gamma_{a\dot{a}}^i s_L^{\dot{a}}
+R\leftrightarrow L
\right)
\ .
\end{eqnarray}
Other known terms do not contribute to the amplitude as they are not
$(1,1)$ operators.
In the same reasoning, matrix model vertex operator (\ref{vodilatino}) 
does not contribute to the amplitude.
Although there remain undetermined terms in dilatino vertex operators in
IIB matrix model, we believe that complete set of the operators is 
equivalent to those in type IIA superstring theory.

\section{Conclusion \label{6}}

We have constructed type IIA closed string vertex operators
directly from IIB matrix model on the two dimensional 
noncommutative backgrounds.
The vertex operators which couple to supergravity multiplet were
determined up to the six-th order of Majorana-Weyl spinor $\lambda$
in IIB matrix model \cite{KMS}.
In our analysis,
gravitino, graviton, the fourth rank antisymmetric tensor field and
the second rank antisymmetric tensor fields
which contains up to 6 $\lambda$'s,
show the 
perfect agreement with the corresponding superstring vertex operators. 
In this comparison, 
the identification of noncommutative scale $\frac{1}{\theta}$ 
with string scale 
$\alpha'$ has played an important role, which we have adopted
in the process of deriving the action \cite{KN3}.
Originally, the scale in NC gauge theory
is identified with string scale in the dual
supergravity description \cite{IIKK2,KTT2,KN2}.
Since we take the commutative limit in our formulation,
the direct relation between UV finiteness of string scattering amplitude 
and the regularization of UV divergence which give rise to UV/IR mixing effect 
in noncommutative gauge theory is unclear.
It is interesting to investigate the relation between them in our formulation.
The results in this paper and the previous paper \cite{KN3}
are summarized in Figure 1.
The process (1)+(2) is described in \cite{KN3}.
In this paper, the relation (4) is explicitly 
demonstrated, which can be regarded as the confirmation of the 
other process, especially process (3).
At the first sight, on two dimensional backgrounds of IIB 
matrix model,
the vertex operators are extremely complicated.
But supersymmetry
restricts the possible terms and the operators relevant
to the amplitudes become the same as the vertex operators in 
superstring theory.
After identifying the physical states, we can calculate the multi-point 
scattering amplitude in a standard way.

We have not yet reproduced the complete
vertex operators in IIB matrix model due to algebraic complexity.
Since the 
complete forms of the conjugate 
dilatino and dilaton vertex operators are not yet known,
we can not compare these operators.
However, 
we have found all the necessary 
pieces of each GS superstring vertex operator in 
IIB matrix model vertex operators even in those cases.

\begin{figure}[hbt]
\begin{center}
\includegraphics[height=2cm]{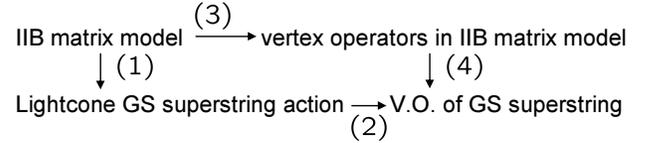} 
\caption{Vertex operators for Green-Schwarz light-cone superstring are
derived from IIB matrix model. The constructions (1)+(2) and (3)+(4)
 should give the same result.}
\end{center}
\end{figure}

\begin{center} \begin{large}
Acknowledgments
\end{large} \end{center}
This work is supported in part by the Grant-in-Aid for Scientific
Research from the Ministry of Education, Science and Culture of Japan.
The work of S.N. is supported in part by the Research Fellowship of
the Japan Society for the Promotion of Science for Young Scientists.

\appendix

\section{Open superstring vertex operators}

The bosonic (vector) and fermionic (spinor) vertex operators 
of light-cone open superstring are written 
as
\begin{eqnarray}
V_B (\zeta, k)&=&\zeta^\mu V_\mu^B (k)=
(\zeta^i B^i -\zeta^- B^+)e^{ik \cdot \phi} \ , \n 
V_F (u,k)&=&u^a V_{1a}^F (k)+u^{\dot{a}} V_{2\dot{a}}^F(k)
= (u^a F_1^a +u^{\dot{a}} F_2^{\dot{a}})e^{ik \cdot \phi} \ .
\n
\label{openvertex}
\end{eqnarray}
where $B^i$, $B^+$, $F_1^a$ and $F_2^{\dot{a}}$ are written 
in terms of $\phi_i$ and $s^a$ as
\begin{eqnarray}
B^+&=&p^+ \ , \n 
B^i&=& \left(\dot{\phi}^i -R_{ij} k^j \right) \ , \n 
F_2^{\dot{a}}&=&
(2p^+)^{-1/2} \left[(\Gamma \cdot \dot{\phi} s)^{\dot{a}}+
\frac{1}{3} (\Gamma^i s)^{\dot{a}} R^{ij} k^j \right] \ , \n
F_1^{a}&=&  (\frac{p^+}{2})^{1/2} s^a \ .
\end{eqnarray} 
$s^a$ belong to $8_s$ representation 
in our convention.
$R^{ij}(\tau)$ is defined by
\begin{eqnarray}
R^{ij}(\tau )=  \frac{1}{4} \Gamma_{ab}^{ij}s^a (\tau) s^b (\tau) \ ,
\end{eqnarray}
where
\begin{eqnarray}
\Gamma_{ab}^{ij} \equiv \frac{1}{2} (\gamma_{a \dot{a}}^i
\gamma_{\dot{a} b}^j-\gamma_{a \dot{a}}^j \gamma_{\dot{a} b}^i) \ .
\end{eqnarray}
The matrices $\Gamma^i$ are represented in the 16-dimensional $(8_s+8_c)$
representation of spin (8) as
\begin{eqnarray}
\Gamma^i = 
\left(\begin{array}{ccc}
0&\gamma_{a \dot{a}}^i \\
\gamma_{\dot{b} b}^i&0
\end{array}\right) \ .
\end{eqnarray}
We consider the operators which carry the momentum $k^\mu$ with 
$k^+=0, (k^i)^2=0$.
$(\zeta^+,\zeta^-,\zeta^i)$ represents 
the wave function for the vector state, and $(u^a , u^{\dot{a}})$
represents the wave function for the spinor state.

\section{Type IIA closed string states in type IIB matrix model
\label{4}}
\setcounter{equation}{0}

In order to calculate the multi point superstring amplitude,
we have to identify closed string states.
Closed string states are constructed by the direct products of the left movers 
and right movers in string theory. 
In IIB matrix model on the two dimensional background, 
we can also construct closed string states
in a radial quantization 
as the product of the separate states
corresponding to the left movers and right movers, respectively.

Since the origin in the coordinate system $z$ becomes the infinite past
in the conformal mapping (\ref{conformal}),
we can insert a local operator at the origin and
obtain its charges by the appropriate contour integrals around the origin.
The asymptotic states correspond to the local operators.
\footnote{They become fuzzy at the NC scale.}

Before we define the massless ground states 
of closed strings,
we define the massless ground states of open strings
\begin{eqnarray}
|i \rangle \ , \quad |\dot{a} \rangle \ ,
\end{eqnarray}
as the states in the $8_v(8_c)$ representation of spin (8).
They are normalized as 
\begin{eqnarray}
\langle i | j \rangle =\delta_{ij} \ , \quad 
\langle \dot{a} | \dot{b} \rangle =\delta_{\dot{a}\dot{b}} \ .
\end{eqnarray}

Any physical states $|\Lambda,k \rangle $ are obtained 
by inserting the vertex operators in the far past as
\begin{eqnarray}
|\Lambda , k \rangle =\lim_{\tau \to - \infty} e^{- \tau}
V_B (k) |0,0\rangle \ .
\end{eqnarray}
It is because zero mode operator $Z_0$ acts as
\begin{eqnarray}
Z_0 |0,0 \rangle \equiv 
e^{ik \cdot \phi} z^{k\cdot p+1} |0,0 \rangle =z |0,k \rangle \ , \n
\langle 0,0 | Z_0 \equiv
\langle 0,0 | z^{k \cdot p -1} e^{ik \cdot x }=\frac{1}{z} \langle 0,k |
\ .
\end{eqnarray}
The massless vector states transform as 
\begin{eqnarray} \label{helicityop}
R_0^{ij}|k \rangle &\equiv& \frac{1}{4} s_0^a \Gamma_{ab}^{ij} s_0^b 
|k \rangle \n 
&=&\delta^{jk} |i \rangle -\delta^{ik} |j \rangle \ .
\end{eqnarray}
where $s_0$ is the zero mode of $s$.
$R_0$ is the zero mode helicity operator.
The massless spinor states transform as
\begin{eqnarray}
R_0^{ij} |\dot{a} \rangle =-\frac{1}{2} \Gamma_{\dot{a}\dot{b}}^{ij}
|b \rangle \ .
\end{eqnarray} 
The ground states are mapped to each other by the fermionic zero mode as
\begin{eqnarray}
s_0^a | \dot{a} \rangle &=& \frac{1}{\sqrt{2}} \Gamma_{a \dot{a}}^i |i
 \rangle \ , \n
s_0^a | i \rangle &=& \frac{1}{\sqrt{2}} \Gamma_{a \dot{a}}^i |\dot{a}
 \rangle \ .
\end{eqnarray}
A vector state $|\zeta \rangle$ is defined by
\begin{eqnarray}
|\zeta \rangle = |i \rangle \zeta^i \ ,
\end{eqnarray}
and a spinor state $|u \rangle$ is defined by
\begin{eqnarray}
|u \rangle = |\dot{a} \rangle \frac{u^{\dot{a}} (k)}{\sqrt{k^+}} \ .
\end{eqnarray}
Closed string states are constructed by the direct product of the 
left-movers and 
right-movers.

\end{document}